\documentclass[conference,a4paper]{authorver}
\usepackage{multirow}
\usepackage[dvipdfmx]{graphicx}
\usepackage{amsmath}
\usepackage{bbold}
\usepackage[psamsfonts]{amssymb}
\usepackage{amsxtra}
\usepackage{threeparttable}
\def\bm#1{\mbox{\boldmath $#1$}}
\def\sim#1{s(#1)}

\def\figA{
\begin{figure*}[t]
\begin{center}
\includegraphics[width=180mm]{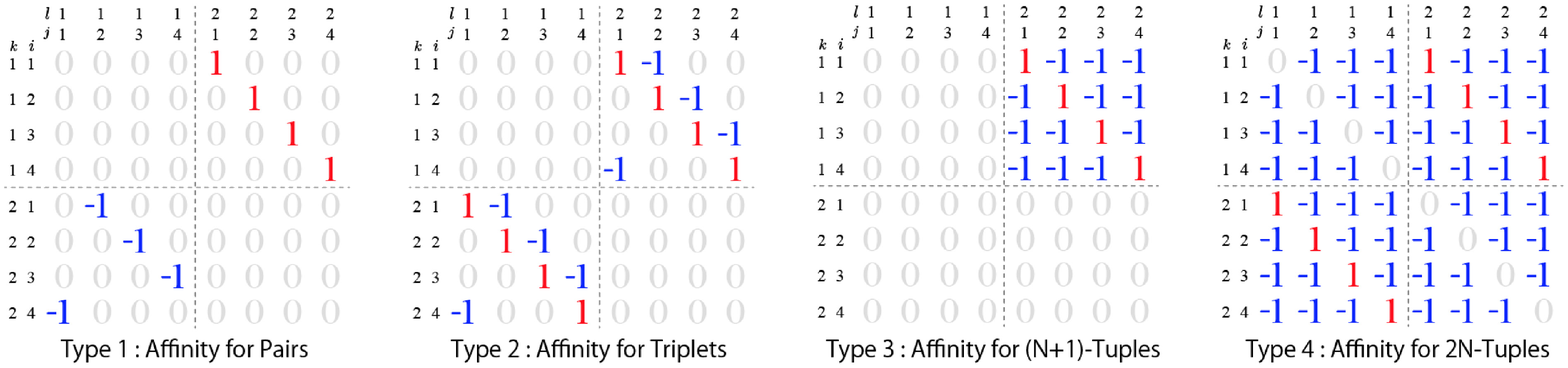}
\end{center}
\vspace*{-10pt}
\caption{
Four types of the affinity tensor $\alpha_{ij}^{kl}$.
The values $1$ and $-1$ denote
representation pairs predisposed to be 
close to and far from each other, respectively.
The diagonal $0$ values are for anchors,
and the other $0$ values make no restriction on sample pairs.
}
\label{figA}
\vspace*{-3pt}
\end{figure*}
}
\def\figB{
\begin{figure}[t]
\begin{center}
\includegraphics[width=80mm]{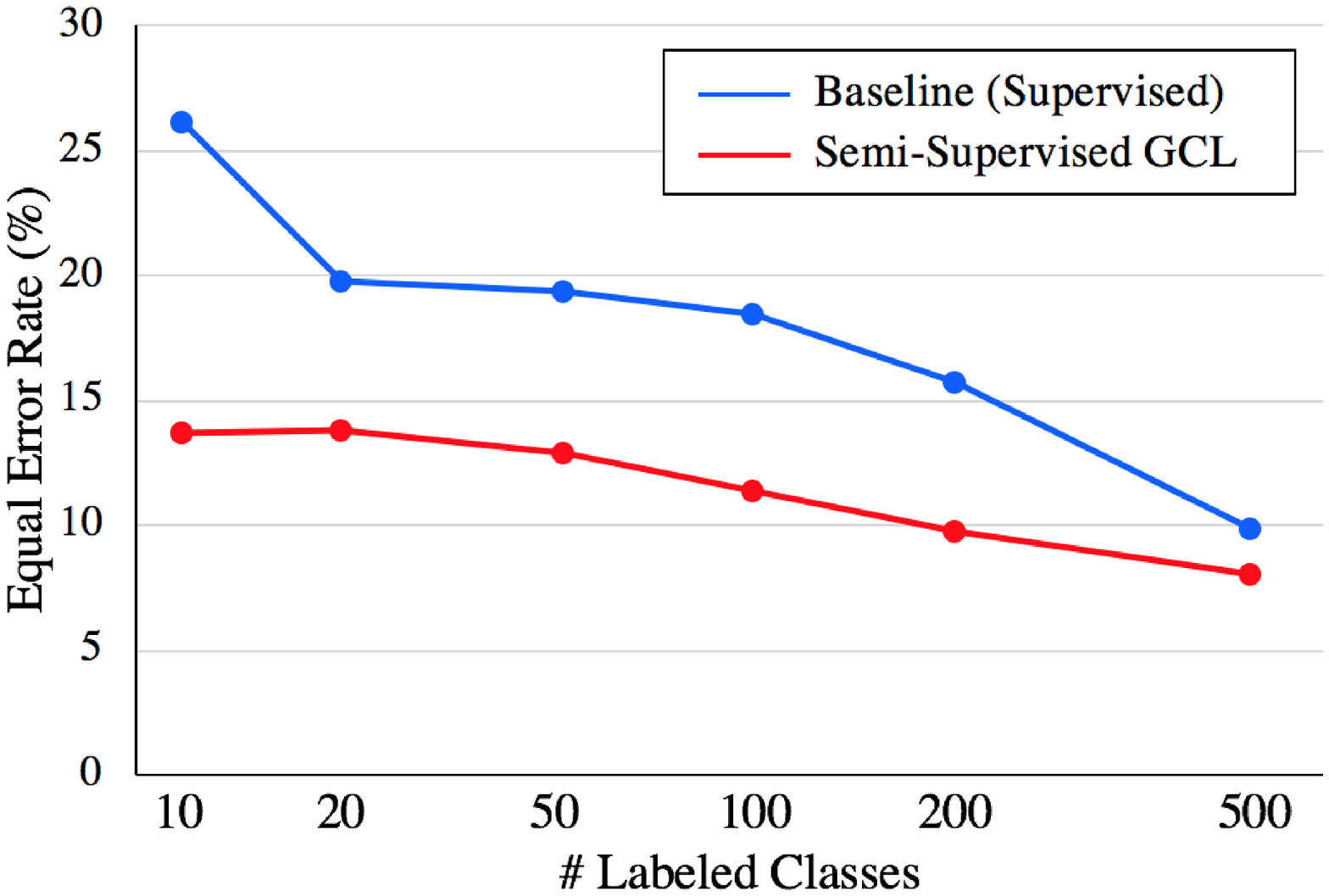}
\end{center}
\vspace*{-10pt}
\caption{
Results for semi-supervised experiments.
The equal error rate on the VoxCeleb 1 test set is reported.
The baseline uses only labeled samples.
Semi-supervised GCL uses both labeled and unlabeled samples.
}
\label{figB}
\vspace*{-3pt}
\end{figure} }
\def\figC{
\begin{figure}[t]
\begin{center}
\includegraphics[width=70mm]{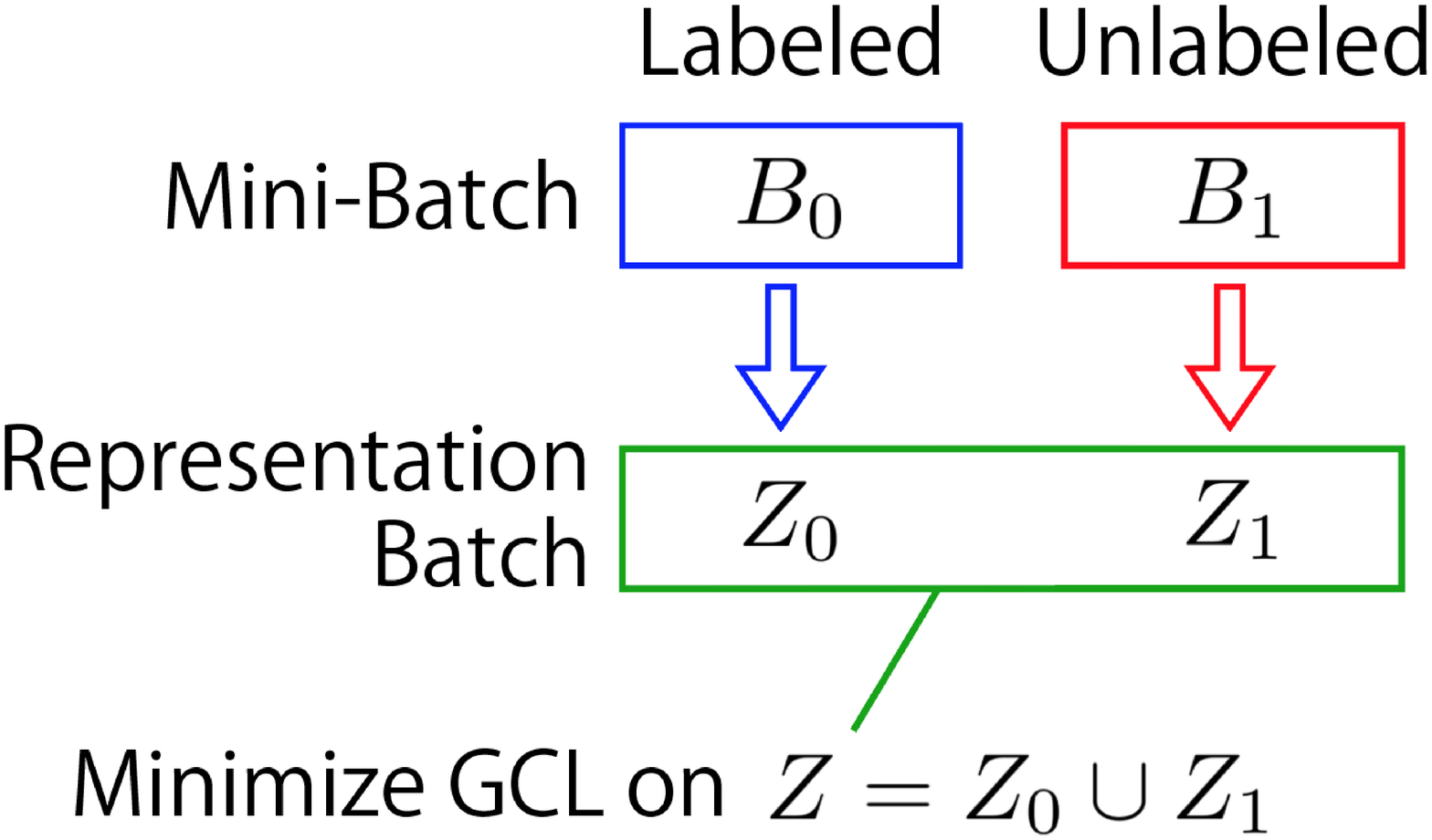}
\end{center}
\vspace*{-10pt}
\caption{
Semi-supervised learning using generalized contrastive loss (GCL).
From a given mini-batch $(B_{0},B_{1})$, which includes both labeled and unlabeled samples, a representation batch $Z=Z_{0} \cup Z_{1}$ is constructed.
$Z_{0}$ is constructed in the same way as in supervised metric learning, for example, with anchors and prototypes.
$Z_{1}$ is constructed in the same way as in unsupervised contrastive learning, for example, with data augmentation functions.
}
\label{figC}
\vspace*{-3pt}
\end{figure}
}

\def\tableA{
\begin{table}[t]
\begin{center}
\caption{Results of semi-supervised, unsupervised, and supervised learning. Equal error rate (EER) on the VoxCeleb 1 test is reported.\label{tableA}}
\begin{tabular}{l|l|l|r}
\hspace{-1.0pt}Method &\hspace{-3pt}Training\hspace{1pt}Scenario \hspace{-7pt}\ &\hspace{-3pt}Additional\hspace{1pt}Data/Model\hspace{-3pt}\ &\hspace{-3pt}EER\hspace{1pt}(\%)\\
\hline
\hspace{-1.0pt}SSL embedding \cite{stafylakis2019self} \hspace{-7pt}\  &\hspace{-3pt}Semi-supervised &\hspace{-3pt}Speech recognition & 6.31\\
\hspace{-1.0pt}Ours &\hspace{-3pt}Semi-supervised &\hspace{-3pt}- & {\bf 6.01}\\
\hline
\hspace{-1.0pt}Cross-modal \cite{nagrani2020disentangled} &\hspace{-3pt}Unsupervised &\hspace{-3pt}Video (face images) & 20.09\\
\hspace{-1.0pt}Ours &\hspace{-3pt}Unsupervised &\hspace{-3pt}- & {\bf 15.26}\\
\hline
\hspace{-1.0pt}AM-Softmax &\hspace{-3pt}Supervised &\hspace{-3pt}- & {\bf 1.81}\\
\hspace{-1.0pt}Ours &\hspace{-3pt}Supervised &\hspace{-3pt}- & 2.56\\
\end{tabular}
\end{center}
\end{table}
}
\def\tableB{
\begin{table*}[t]
\begin{center}
\caption{
Comparison of recent loss definitions in GCL formulation. The affinity tensor makes pairs, triplets, $(N+1)$-tuples, or $2N$-tuples, as shown in Figure~\ref{figA}.
Representation batch $Z$ is constructed from labeled samples, unlabeled samples, and/or parameters.
See the definition of GCL in Sec.\ IV for the meaning of $s$, $\tilde \alpha$, and $\Psi(v)$.
$m$ is a margin hyper-parameter, and $M=\sum_{j,l} s(\bm{z}_{i}^{k},\bm{z}_{j}^{l};\alpha_{ij}^{kl})+\epsilon$.
\label{tableB}
}
\begin{tabular}{l|c|c|c|c|c}
Loss & Affinity & Representation Batch $Z$ & Similarity $s(\bm{z},\bm{z}';\alpha)$ & $\tilde \alpha$ & $\Psi(v)$\\
\hline
Contrastive loss \cite{hadsell2006dimensionality}& Type 1 & Labeled & $\alpha d(\bm{z},\bm{z}')$ & $\alpha$ & $-(\langle v \rangle - \chi(v<0) \langle v + m \rangle)/2$\\
Triplet loss \cite{hoffer2015deep}& Type 2 & Labeled & $\alpha d(\bm{z},\bm{z}')$ & $\alpha$ & $- \langle v+m \rangle$\\
ArcFace (AAM loss) \cite{deng2019arcface}& Type 3 & Labeled+weights &
$|\alpha| \exp(\cos(\angle (\bm{z},\bm{z}')+m\langle \alpha \rangle))$ & $\langle \alpha \rangle$ & $2v/M$\\
SphereFace \cite{liu2017sphereface}& Type 3 & Labeled+weights &
$|\alpha| \exp((1+m \langle \alpha \rangle) \cos(\bm{z},\bm{z}'))$ & $\langle \alpha \rangle$ & $2v/M$\\
CosFace \cite{wang2018cosface}& Type 3 & Labeled+weights &
$|\alpha| \exp(\cos(\bm{z},\bm{z}')-m\langle \alpha \rangle)$ & $\langle \alpha \rangle$ & $2v/M$\\
Prototypical episode loss \cite{snell2017prototypical}& Type 3 & Labeled &
$|\alpha| \exp(-d(\bm{z},\bm{z}'))$ & $\langle \alpha \rangle$ & $2v/M$\\
Angle-prototypical loss \cite{chung2020in}& Type 3 & Labeled &
$|\alpha| \exp(\gamma \cos(\bm{z},\bm{z}') + \beta)$ & $\langle \alpha \rangle$ & $2v/M$\\
SimCLR (NT-Xent loss) \cite{chen2020simple}& Type 4 & Unlabeled &
$|\alpha| \exp(\cos(g(\bm{z}),g(\bm{z}'))/\tau)$ & $\langle \alpha \rangle$ & $v/M$\\
Our experimental setting & Type 4 & Labeled+unlabeled &
$|\alpha| \exp(\gamma \cos(\bm{z},\bm{z}') + \beta)$  & $\langle \alpha \rangle$ & $v/M$ 
\end{tabular}
\end{center}
\end{table*}
}

\begin{document}

\title{
Semi-Supervised Contrastive Learning with Generalized Contrastive Loss
and\\Its Application to Speaker Recognition} 

\author{%
\authorblockN{%
Nakamasa Inoue and Keita Goto
}
\authorblockA{%
Tokyo Institute of Technology, Tokyo, Japan\\
E-mail: inoue@c.titech.ac.jp}
}

\maketitle
\thispagestyle{empty}

\begin{abstract}
This paper introduces a semi-supervised contrastive learning framework and its application to text-independent speaker verification. The proposed framework employs generalized contrastive loss (GCL).
GCL unifies losses from two different learning frameworks, supervised metric learning and unsupervised contrastive learning, and thus it naturally determines the loss for semi-supervised learning.
In experiments, we applied the proposed framework to text-independent speaker verification on the VoxCeleb dataset. We demonstrate that GCL enables the learning of speaker embeddings in three manners, supervised learning, semi-supervised learning, and unsupervised learning, without any changes in the definition of the loss function.
\end{abstract}

\section{Introduction}

With the development of various optimization techniques, deep learning has become a powerful tool for numerous applications, including speech and image recognition.
To build high-performance models, supervised learning is the most popular methodology, in which labeled samples are used for optimizing model parameters.
It is known that deep neural networks (e.g., ResNet \cite{he2016deep}) having more than a million parameters outperform hand-crafted feature extraction methods.
As such, optimizing parameters with a well-designed objective function is one of the most important research topics in deep learning.

In recent years, supervised metric learning methods for deep neural networks have attracted attention. Examples of these include triplet loss \cite{hoffer2015deep} and prototypical episode loss \cite{snell2017prototypical}, which predispose a network to minimize within-class distance and maximize between-class distance.
They are also effective for text-independent speaker verification, as shown in \cite{chung2020in}, because cosine similarity between utterances from the same speaker is directly maximized in the training phase.

Nevertheless, unsupervised learning methods have grown greatly, thanks to large-scale collections of unlabeled samples. Some studies have recently proven that self-supervised learning achieves performance very close to that of supervised learning. For example, the simple framework for contrastive learning of representations (SimCLR) \cite{chen2020simple} provides superior image representation by introducing contrastive NT-Xent loss using data augmentation on unlabeled images. For speaker verification, these methods motivate us to explore unsupervised and semi-supervised ways to learn speaker embeddings by effectively using unlabeled utterances.

In general, supervised learning and unsupervised learning depend on different methodologies. However, supervised metric learning and unsupervised contrastive learning share a common idea to maximize or minimize the similarity between samples.
This implies the possibility of unifying these two learning frameworks.

In this paper, we propose a semi-supervised contrastive learning framework based on generalized contrastive loss (GCL).
GCL provides a unified formulation of two different losses from supervised metric learning and unsupervised contrastive learning.
Thus, it naturally works as a loss function for semi-supervised learning.
In experiments, we applied the proposed framework to text-independent speaker verification on the VoxCeleb dataset. We demonstrated that GCL enables the network to learn speaker embeddings in three manners, supervised learning, semi-supervised learning, and unsupervised learning, without any changes in the definition of the loss function.

\section{Related Work}
\subsection{Supervised Metric Learning}
\label{sec2.1}
Supervised metric learning is a framework to learn a metric space from a given set of labeled training samples.
For recognition problems, such as audio and image recognition,
the goal is typically to learn the semantic distance between samples.

A recent trend in supervised metric learning is to
design a loss function at the top of a deep neural network.
Examples include
contrastive loss for Siamese networks \cite{hadsell2006dimensionality},
triplet loss for triplet networks \cite{hoffer2015deep},
and episode loss for 
prototypical networks \cite{snell2017prototypical}.
To measure the distance between samples, 
Euclidean distance is often used with these losses.

For face identification from images, 
measuring similarity by cosine similarity often improves the performance.
ArcFace \cite{deng2019arcface}, CosFace \cite{wang2018cosface}, and SphereFace \cite{liu2017sphereface} are its popular implementations.
Their effectiveness is also shown in speaker verification from audio samples
with some extended loss definitions, such as ring loss \cite{zheng2018ring, Liu2019ring}.
One of the best choices for speaker verification is angle-prototypical loss \cite{chung2020in}, which introduces cosine similarity to episode loss, as shown in \cite{chung2020in} with thorough experiments. 

\subsection{Unsupervised Contrastive Learning}
\label{sec2.2}
Unsupervised learning is a framework to train
a model from a given set of unlabeled training samples.
Classic methods for unsupervised learning include
clustering methods such as $K$-means clustering \cite{lloyd1982least}.
Most of them are statistical approaches with some objectives based on means and variances.

Recently, self-supervised learning has proven to be effective
for pre-training deep neural networks.
For example, Jigsaw \cite{noroozi2016unsupervised} and Rotation \cite{gidaris2018unsupervised} define a pretext task on unlabeled data and pre-train networks for image recognition by solving it.
Deep InfoMax \cite{hjelm2018learning} and its multiscale extension AMDIM \cite{bachman2019learning} focus on mutual information between representations extracted from multiple views of a context.
SimCLR \cite{chen2020simple} introduces contrastive learning using data augmentation.
The effectiveness of contrastive learning is also shown in MoCo V2 \cite{he2019momentum, chen2020improved}.
These methods achieve performance comparable with that of supervised learning in tasks of image representation learning.

Cross-modal approaches are also effective if more than one source is available.
For speaker verification, Nagrani et al.\ \cite{nagrani2020disentangled} proposed a cross-modal self-supervised learning method, which uses face images as supervision of audio signals to identify speakers.

\subsection{Semi-Supervised Learning}
Semi-supervised learning is a framework to 
train a model from a set consisting of both labeled and unlabeled samples.
To effectively incorporate information from unlabeled samples into the parameter optimization step, a regularization term is often introduced into the objective function.
For example, consistency regularization \cite{sajjadi2016regularization} is used to penalize sensitivity to augmented unlabeled samples.

For speaker verification, Stafylakis et al.\ \cite{stafylakis2019self} proposed self-supervised speaker embeddings. A pre-trained automatic speech recognition system is utilized to make a supervision signal of phoneme information on unlabeled utterances.

\section{Preliminary} \subsection{Supervised Metric Learning}
\label{sec3.1}
Let $\mathcal{D}$ be a training dataset for supervised learning,
which consists of sample pairs $\bm{x}$ and their discrete class label $y$.
The goal of supervised metric learning is to learn a metric function $d(\bm{x},\bm{x}')$,
which assigns a small distance between samples belonging to the same class,
and relatively large distance between samples from different classes. 
Assuming that the training phase has iterations for parameter updates,
a mini-batch $B$ is sampled from $\mathcal{D}$ at each iteration.
For convenience, two-step sampling is often used \cite{chung2020in}.
First, a set of $N$ different classes are randomly sampled from the set of training classes.
We denote the sampled classes by $y_{1},y_{2},\cdots,y_{N}$.
Second, $K$ independent samples are randomly sampled from each of $N$ classes.
We denote the samples from the class $y_{i}$ as $\bm{x}_{i}^{1}, \bm{x}_{i}^{2}, \cdots, \bm{x}_{i}^{k}$.
As a result, a mini-batch
$B = \{(\bm{x}_{i}^{k},y_{i}):i=1,2,\cdots,N, k=1,2,\cdots,K\}$
consists of $NK$ samples.

As an example of supervised metric learning,
we show the training process of a prototypical network \cite{snell2017prototypical}.
The main idea of a prototypical network is to make prototype representations of each class and to minimize the distance between a query sample and its corresponding prototype. Its loss for parameter updates is computed as follows:
\begin{enumerate}
\item Sample a mini-batch  $B$ from $\mathcal{D}$ and split it into a query set
$Q=\{(\bm{x}_{i}^{1},y_{i})\in B:k=1\}$
and a support set $S=\{(\bm{x}_{i}^{k},y_{i})\in B:k>1\}$.
\item Extract query representations $\bm{z}_{i}^{1}$ from $Q$ by
\begin{align}
\bm{z}_{i}^{1} = f_{\theta}(\bm{x}_{i}^{1}),
\end{align}
where $f_{\theta}$ is a neural network for embedding (i.e., a network without the final loss layer) and $\theta$ is a set of parameters.
\item Construct prototype representations $\bm{z}_{i}^{2}$ from $S$ by
\begin{align}
\bm{z}_{i}^{2} = \frac{1}{K-1} \sum_{k=2}^{K} f_{\theta}(\bm{x}_{i}^{k}).
\end{align}
\item
From a representation batch $Z = \{\bm{z}_{i}^{k} : i=1,2,\cdots,N, k=1,2\}$, compute the episode loss defined by
\begin{align}
\label{mt}
L = - \frac{1}{N} \sum_{i} \frac{\sim{\bm{z}_{i}^{1}, \bm{z}_{i}^{2}}}{\sum_{j}
\sim{\bm{z}_{i}^{1}, \bm{z}_{j}^{2}}},
\end{align}
where $s$ is the exponential function of negative distance between representations
$\sim{\bm{z}, \bm{z}'} := \exp(- d(\bm{z}, \bm{z}'))$,
and $d$ is the squared Euclidean distance.
\end{enumerate}

\subsection{Unsupervised Contrastive Learning}
\label{sec3.2}
Let $\mathcal{U}$ be a training dataset for unsupervised learning,
which consists of unlabeled samples $\bm{u}$.
The goal of unsupervised learning is to train networks without any manually attached labels.

As an example of unsupervised learning, we show the training process of SimCLR \cite{chen2020simple}.
SimCLR maximizes the similarity between representations of two augmented samples $t_{1}(\bm{u})$ and $t_{2}(\bm{u})$,
where $t_{1}$ and $t_{2}$ are two randomly selected augmentation functions.
Its loss for parameter updates is computed as follows:

\begin{enumerate}
\item Sample a mini-batch $B = \{\bm{u}_{i}:i=1,2,\cdots,N\}$ from $\mathcal{U}$.
\item Extract the first representation $\bm{z}_{i}^{1}$ by
\begin{align}
\bm{z}_{i}^{1} = f_{\theta}(t_{1}(\bm{u}_{i})).
\end{align}
Note that $t_{1}$ is randomly selected from a set of augmentation functions for each $i$.
\item Extract the second representation $\bm{z}_{i}^{2}$ by
\begin{align}
\bm{z}_{i}^{2} = f_{\theta}(t_{2}(\bm{u}_{i})).
\end{align}
\item 
From a representation batch $Z = \{\bm{z}_{i}^{k} : i=1,2,\cdots,N, k=1,2\}$, compute the NT-Xent loss \cite{chen2020simple} defined by
\begin{align}
L_{s} = \frac{1}{2} (\ell_{12} + \ell_{21}),
\end{align}
where
\begin{align}
\label{ell12}
\ell_{12} = - \frac{1}{N} \sum_{i} \frac{\sim{\bm{z}_{i}^{1}, \bm{z}_{i}^{2}}}{\sum_{j} \sim{\bm{z}_{i}^{1}, \bm{z}_{j}^{2}} + \sum_{j \not = i} \sim{\bm{z}_{i}^{1}, \bm{z}_{j}^{1}}
},\\
\label{ell21}
\ell_{21} = - \frac{1}{N} \sum_{i} \frac{\sim{\bm{z}_{i}^{2}, \bm{z}_{i}^{1}}}{\sum_{j} \sim{\bm{z}_{i}^{2}, \bm{z}_{j}^{1}} + \sum_{j \not = i} \sim{\bm{z}_{i}^{2}, \bm{z}_{j}^{2}}.
}
\end{align}
Here $s$ is the exponential of similarity between representations $\sim{\bm{z},\bm{z}'}=\exp(\cos(g_{\theta'}(\bm{z}),g_{\theta'}(\bm{z}'))/\tau)$, $g_{\theta'}$ is a fully connected layer with a parameter $\theta'$, and $\tau$ is a hyper-parameter.
\end{enumerate}
We note that by omitting the second summation in the denominator of Eq.~(\ref{ell12}) or (\ref{ell21}) we obtain Eq.~(\ref{mt}).
This opens a way to bridge the two losses for supervised metric learning and unsupervised contrastive learning.

\section{Proposed Method}
This section presents
1) Generalized contrastive loss (GCL) and
2) GCL for semi-supervised learning.
GCL unifies losses from two different learning frameworks, supervised metric learning and unsupervised contrastive learning, and thus it naturally works as a loss function for semi-supervised learning.

\subsection{Generalized Contrastive Loss}
Let $Z = \{\bm{z}_{i}^{k} : i=1,2,\cdots,N, k=1,2\}$
be a representation batch obtained from a mini-batch for either supervised metric learning or unsupervised contrastive learning (see Step 4 in Sec.~\ref{sec3.1} and Sec.~\ref{sec3.2}).
We define the GCL as
\begin{align}
L_{\alpha} = \frac{1}{2N} \sum_{i,k}
\frac{
\sum_{j,l} \langle \alpha_{ij}^{kl} \rangle \sim{\bm{z}_{i}^{k}, \bm{z}_{j}^{l}}}{
\sum_{j,l} |\alpha_{ij}^{kl}| \sim{\bm{z}_{i}^{k}, \bm{z}_{j}^{l}}+\epsilon},
\end{align}
where $\alpha_{ij}^{kl}$ is a fourth-order affinity tensor, $\langle \cdot \rangle$ denotes the application of Macaulay brackets to the ramp function as $\langle a \rangle = \max(0, a)$, and $\epsilon \simeq 0$ is a constant to avoid a division by zero.
Note that a positive value for $\alpha_{ij}^{kl}$ predisposes
two representations $\bm{z}_{i}^{k}$ and $\bm{z}_{j}^{l}$ to be close to each other,
a negative value does the opposite.
The episode loss can be viewed as a special case of GCL
when $Z$ is made from a mini-batch of labeled samples via prototypes, as shown in Sec.~\ref{sec3.1}, and the affinity tensor is defined by
\begin{align}
\alpha_{ij}^{kl} =
\begin{cases}
1 & (k < l, i = j)\\
-1 & (k < l, i \not = j)\\
0 & (\mbox{otherwise})
\end{cases}.
\end{align}
Note that $i$ is the category index and $k$ is the sample index in this case.

\figC

The NT-Xent loss can also be viewed as a special case of GCL
when $Z$ is made from a mini-batch of unlabeled samples using augmentation, as shown in Sec.~\ref{sec3.2}, and the affinity tensor is defined by
\begin{align}
\alpha_{ij}^{kl} =
\begin{cases}
1 & (k \not = l, i = j)\\
0 & (k = l, i = j)\\
-1 & (\mbox{otherwise})
\end{cases}.
\end{align}
Note that $i$ is the sample index and $k$ is the augmentation type index in this case.

Other types of losses, including generalized end-to-end loss \cite{wan2018generalized} and
angle-prototypical loss \cite{chung2020in}, can also be obtained by
changing the definitions of $Z$, $\alpha$, and $s$.
Note that the complete definition of GCL includes more instances
of metric learning methods, as discussed in the Appendix.

\subsection{GCL for Semi-Supervised Learning}
In semi-supervised learning, a training dataset includes both labeled and unlabeled samples.
Thus, a mini-batch is given by a pair $B=(B_{0}, B_{1})$ of
a set of labeled samples $B_{0}$ and a set of unlabeled samples $B_{1}$.
To apply GCL to $B$,
its representation batch is constructed by $Z = Z_{0} \cup Z_{1}$, where
\begin{align}
Z_{0} &= \{\bm{z}_{i|0}^{k} : i=1,2,\cdots,N, k=1,2\}
\end{align}
is a representation batch of $B_{0}$ given from a supervised metric learning method and
\begin{align}
Z_{1} &= \{\bm{z}_{i|1}^{k} : i=1,2,\cdots,N', k=1,2\}
\end{align}
is a representation batch of $B_{1}$ given from an unsupervised contrastive learning method, as shown in Figure~\ref{figC}.

The GCL for semi-supervised learning is then defined on $Z$ by
\begin{align}
L_{\alpha} = \sum_{i,k,u}
\frac{
\sum_{j,l,v} \langle \alpha_{ij|uv}^{kl} \rangle \sim{\bm{z}_{i|u}^{k}, \bm{z}_{j|v}^{l}}}{
\sum_{j,l,v} |\alpha_{ij|uv}^{kl}| \sim{\bm{z}_{i|u}^{k}, \bm{z}_{j|v}^{l}},
}
\end{align}
where $u, v \in \{0,1\}$ denote labeled or unlabeled samples.
Note that affinity tensor $\alpha_{ij|uv}^{kl}$ becomes a sixth-order tensor
to predispose similarity between $\bm{z}_{i|u}^{k}$ and $\bm{z}_{j|v}^{l}$ to be close or far.

Here, we provide an example definition of $\alpha_{ij|uv}^{kl}$ for semi-supervised learning.
Compared with NT-Xent loss, we relax the affinity between unlabeled samples
because some labeled samples are available for training.
\begin{align}
\alpha_{ij|00}^{kl} &=
\begin{cases}
1 & (k \not = l, i = j)\\
0 & (k = l, i = j)\\
-1 & (\mbox{otherwise})
\end{cases}\\
\alpha_{ij|11}^{kl} &=
\begin{cases}
1 & (k \not = l, i = j)\\
0 & (k = l, i = j)\\
-1 & (\mbox{otherwise})
\end{cases}\\
\alpha_{ij|01}^{kl} &= -1\\
\alpha_{ij|10}^{kl} &= -1.
\end{align}
This definition is effective for semi-supervised learning for speaker verification,
where labeled utterances are from a pre-defined set of speakers
and unlabeled utterances are from another (different) set of unknown speakers. For the similarity measure, we use $\sim{\bm{z},\bm{z}'}=\exp(\gamma \cos(\bm{z},\bm{z}')+\beta)$. This definition is used in \cite{chung2020in}.

\section{Experiments}
\subsection{Experimental Settings}
We used the VoxCeleb dataset \cite{nagrani2017voxceleb,chung2018voxceleb} for evaluating our proposed framework.
The training set (voxceleb\_2\_dev) consists of 1,092,009 utterances of 5,994 speakers.
The test set (voxceleb\_1\_test) consists of 37,611 enrollment-test utterance pairs.
The equal error rate (EER) was used as an evaluation measure.

For semi-supervised learning experiments,
we randomly selected $P$ speakers from the set of 5,994.
We used their labeled samples and the remaining unlabeled samples for training.
This is the same evaluation setting proposed in \cite{stafylakis2019self}.
For unsupervised learning experiments, we did not use speaker labels.
This evaluation setting is more difficult than the cross-modal self-supervised setting in \cite{ding2020autospeech} because we did not use videos (face images) for training.
For supervised learning experiments, we used all labeled samples  for training.
This is the official evaluation setting on the VoxCeleb dataset.

We used the ResNet18 convolutional network with an input of 40-dimensional filter bank features.
For data augmentation to construct a representation batch from unlabeled samples,
we used four Kaldi data augmentation schemes with the MUSAN (noise, music, and babble) and the RIR (room impulse response) datasets.
For semi-supervised learning, 10~\% of samples in each mini-batch were unlabeled and the others were labeled.

\tableA
\figB

\subsection{Results}
Table~\ref{tableA} summarizes EERs for semi-supervised, unsupervised, and supervised learning settings.
The results demonstrate that GCL enables the learning of speaker embeddings in the three different settings without any changes in the definition of the loss function.

For semi-supervised learning experiments,
we compared the results with those of \cite{stafylakis2019self} by using the same number of labeled speakers ($P=899$).
The results show that our framework achieves comparable performance.
Note that the method in \cite{stafylakis2019self} uses an automatic speech recognition model pre-trained on another dataset, but we did not use such pre-trained models.
Comparison with a supervised learning method is shown in Figure~\ref{figB}.
We see that adding unlabeled utterances improved the performance,
in particular when the number of available labeled utterances was small.

For unsupervised learning experiments,
our method outperformed the cross-modal self-supervised method in \cite{nagrani2020disentangled}.
Note that our method did not use any visual information, such as face images, for supervision.
Audio-visual unsupervised learning with our framework is promising as a next step.

For supervised learning experiments, our method achieves a 2.56~\% EER without using data augmentation. However, there is still room for improvement, because training the same network with Softmax and AM-Softmax losses (training with Softmax and fine-tuning with AM-Softmax) achieves a 1.81~\% EER.
Introducing a more effective network structure, such as ECAPA-TDNN \cite{desplanques2020ecapa} and AutoSpeech-NAS \cite{ding2020autospeech}, to our framework would be also interesting as future work.

\section{Conclusion}

This paper proposed a semi-supervised contrastive learning framework
with GCL.
We showed via experiments on the VoxCeleb dataset that the proposed GCL enables a network to learn speaker embeddings in three manners, namely, supervised learning, semi-supervised learning, and unsupervised learning. Furthermore, this was accomplished without making any changes to the definition of the loss function.

\section*{Acknowledgment}
This work was partially supported by the Japan Science and Technology Agency, ACT-X Grant JPMJAX1905, and the Japan Society for the Promotion of Science, KAKENHI Grant 19K22865.

\bibliographystyle{unsrt}

\begin{thebibliography}{10}

\bibitem{he2016deep}
Kaiming He, Xiangyu Zhang, Shaoqing Ren, and Jian Sun.
\newblock Deep residual learning for image recognition.
\newblock In {\em Proceedings of the International Conference on Computer Vision and Pattern Recognition (CVPR)},
pages 770--778, 2016.

\bibitem{hoffer2015deep}
Elad Hoffer and Nir Ailon.
\newblock Deep metric learning using triplet network.
\newblock In {\em Proceedings of the International Workshop on Similarity-Based Pattern Recognition (SIMBAD)},
pages 84--92, 2015.

\bibitem{snell2017prototypical}
Jake Snell, Kevin Swersky, and Richard Zemel.
\newblock Prototypical networks for few-shot learning.
\newblock In {\em Proceedings of the Advances in Neural Information Processing Systems (NeurIPS)},
pages 4077--4087, 2017.

\bibitem{chung2020in}
Joon~Son Chung, Jaesung Huh, Seongkyu Mun, Minjae Lee, Hee~Soo Heo, Soyeon
  Choe, Chiheon Ham, Sunghwan Jung, Bong-Jin Lee, and Icksang Han.
\newblock In defence of metric learning for speaker recognition.
\newblock {\em arXiv preprint arXiv:2003.11982}, 2020.

\bibitem{chen2020simple}
Ting Chen, Simon Kornblith, Mohammad Norouzi, and Geoffrey Hinton.
\newblock A simple framework for contrastive learning of visual representations.
\newblock {\em Proceedings of the International Conference on Machine Learning (ICML)}, 2020.

\bibitem{hadsell2006dimensionality}
Raia Hadsell, Sumit Chopra, and Yann LeCun.
\newblock Dimensionality reduction by learning an invariant mapping.
\newblock In {\em Proceedings of the International Conference on Computer Vision and Pattern Recognition (CVPR)},
pages 1735--1742, 2006.

\bibitem{deng2019arcface}
Jiankang Deng, Jia Guo, Niannan Xue, and Stefanos Zafeiriou.
\newblock Arcface: Additive angular margin loss for deep face recognition.
\newblock In {\em Proceedings of the International Conference on Computer Vision and Pattern Recognition (CVPR)},
pages 4690--4699, 2019.

\bibitem{wang2018cosface}
Hao Wang, Yitong Wang, Zheng Zhou, Xing Ji, Dihong Gong, Jingchao Zhou, Zhifeng
  Li, and Wei Liu.
\newblock Cosface: Large margin cosine loss for deep face recognition.
\newblock In {\em Proceedings of the International Conference on Computer Vision and Pattern Recognition (CVPR)},
pages 5265--5274, 2018.

\bibitem{liu2017sphereface}
Weiyang Liu, Yandong Wen, Zhiding Yu, Ming Li, Bhiksha Raj, and Le~Song.
\newblock Sphereface: Deep hypersphere embedding for face recognition.
\newblock In {\em Proceedings of the International Conference on Computer Vision and Pattern Recognition (CVPR)},
pages 212--220, 2017.

\bibitem{zheng2018ring}
Yutong Zheng, Dipan~K. Pal, and Marios Savvides.
\newblock Ring loss: Convex feature normalization for face recognition.
\newblock In {\em Proceedings of the International Conference on Computer Vision and Pattern Recognition (CVPR)},
pages 5089--5097, 2018.

\bibitem{Liu2019ring}
Yi Liu, Liang He, and Jia Liu.
\newblock Large Margin Softmax Loss for Speaker Verification.
\newblock In {\em Proceedings of Interspeech},
2019. 

\bibitem{lloyd1982least}
Stuart Lloyd.
\newblock Least squares quantization in pcm.
\newblock {\em IEEE Transactions on Information Theory},
vol.~28, no.~2, pages~129--137, 1982.

\bibitem{noroozi2016unsupervised}
Mehdi Noroozi and Paolo Favaro.
\newblock Unsupervised learning of visual representations by solving jigsaw
  puzzles.
\newblock In {\em Proceedings of the European Conference on Computer Vision (ECCV)},
pages 69--84, 2016.

\bibitem{gidaris2018unsupervised}
Spyros Gidaris, Praveer Singh, and Nikos Komodakis.
\newblock Unsupervised representation learning by predicting image rotations.
\newblock In {\em Proceedings of the International Conference on Learning Representations}, 2018.

\bibitem{hjelm2018learning}
R~Devon Hjelm, Alex Fedorov, Samuel Lavoie-Marchildon, Karan Grewal, Phil
  Bachman, Adam Trischler, and Yoshua Bengio.
\newblock Learning deep representations by mutual information estimation and
  maximization.
\newblock In {\em Proceedings of the International Conference on Learning Representations}, 2019.

\bibitem{bachman2019learning}
Philip Bachman, R.~Devon Hjelm, and William Buchwalter.
\newblock Learning representations by maximizing mutual information across
  views.
\newblock In {\em Proceedings of the Advances in Neural Information Processing Systems (NeurIPS)},
pages 15509--15519, 2019.

\bibitem{he2019momentum}
Kaiming He, Haoqi Fan, Yuxin Wu, Saining Xie, and Ross Girshick.
\newblock Momentum contrast for unsupervised visual representation learning.
\newblock {\em arXiv preprint arXiv:1911.05722}, 2019.

\bibitem{chen2020improved}
Xinlei Chen, Haoqi Fan, Ross Girshick, and Kaiming He.
\newblock Improved baselines with momentum contrastive learning.
\newblock {\em arXiv preprint arXiv:2003.04297}, 2020.

\bibitem{nagrani2020disentangled}
Arsha Nagrani, Joon~Son Chung, Samuel Albanie, and Andrew Zisserman.
\newblock Disentangled speech embeddings using cross-modal self-supervision.
\newblock In {\em Proceedings of the International Conference on Acoustics, Speech and Signal Processing (ICASSP)},
pages 6829--6833, 2020.

\bibitem{sajjadi2016regularization}
Mehdi Sajjadi, Mehran Javanmardi, and Tolga Tasdizen.
\newblock Regularization with stochastic transformations and perturbations for
  deep semi-supervised learning.
\newblock In {\em Proceedings of the Advances in Neural Information Processing Systems (NeurIPS)},
pages 1163--1171, 2016.

\bibitem{stafylakis2019self}
Themos Stafylakis, Johan Rohdin, Oldrich Plchot, Petr Mizera, and Lukas Burget.
\newblock Self-supervised speaker embeddings.
\newblock {\em Proceedings of Interspeech}, 2019.

\bibitem{wan2018generalized}
Li~Wan, Quan Wang, Alan Papir, and Ignacio~Lopez Moreno.
\newblock Generalized end-to-end loss for speaker verification.
\newblock In {\em Proceedings of the International Conference on Acoustics, Speech and Signal Processing (ICASSP)},
pages 4879--4883, 2018.

\bibitem{nagrani2017voxceleb}
Arsha Nagrani, Joon~Son Chung, and Andrew Zisserman.
\newblock Voxceleb: A large-scale speaker identification dataset.
\newblock In {\em Proceedings of Interspeech},
2017.

\bibitem{chung2018voxceleb}
Joon~Son Chung, Arsha Nagrani, and Andrew Zisserman.
\newblock Voxceleb2: Deep speaker recognition.
\newblock In {\em Proceedings of Interspeech},
2018.

\bibitem{ding2020autospeech}
Shaojin Ding, Tianlong Chen, Xinyu Gong, Weiwei Zha, and Zhangyang Wang.
\newblock Autospeech: Neural architecture search for speaker recognition,
\newblock {\em arXiv preprint arXiv:2005.03215}, 2020.

\bibitem{desplanques2020ecapa}
Brecht Desplanques, Jenthe Thienpondt, and Kris Demuynck.
\newblock Ecapa-tdnn: Emphasized channel attention, propagation and aggregation
  in tdnn based speaker verification.
\newblock {\em arXiv preprint arXiv:2005.07143}, 2020.

\end{thebibliography}

\tableB
\figA
\section*{Appendix}
The complete form of the proposed GCL is defined
over a representation batch $Z = \{\bm{z}_{i}^{k} : i=1,2,\cdots,N, k=1,2,\cdots,N\}$ by
\begin{align}
L = - \frac{1}{KN}\sum_{i,k} \Psi
\left(\sum_{j,l} s(\bm{z}_{i}^{k},\bm{z}_{j}^{l};\alpha_{ij}^{kl})\right),
\end{align}
where $\alpha_{ij}^{kl}$ is an affinity tensor,
$s(\bm{z},\bm{z}';\alpha)$ is the similarity between $\bm{z}$ and $\bm{z}'$ given an affinity value $\alpha$,
and $\Psi$ is a normalization or clipping function.

Table~\ref{tableB} summarizes how to obtain popular loss functions from GCL.
We hope this provides an overview of recent progress and helps other researchers
develop new unsupervised, semi-supervised, and supervised learning methods.

\subsection{Affinity Type}
Four types of affinity tensor definitions are used in Table~\ref{tableB}.
With all of them, a positive value for $\alpha_{ij}^{kl}$ predisposes
two representations $\bm{z}_{i}^{k}$ and $\bm{z}_{j}^{l}$ to be close to each other,
a negative value does the opposite. The density of $\alpha_{ij}^{kl}$ increases in the order of Types 1 to 4, as shown in Figure~\ref{figA}.
Definitions of the types are given below. Note that $K=2$ is assumed for simplicity.

{\bf Type 1} makes pairs $(\bm{z}_{i}^{1}, \bm{z}_{j}^{2})$
and its output is $1$ if two samples are from the same class (i.e., $i=j$)
and $-1$ if two samples are from different classes (i.e., $i \not = j$).
An example definition of this type is given by
\begin{align}
\alpha_{ij}^{kl} =
\begin{cases}
1 & (k < l, i = j)\\
-1 & (k > l, i = j-1\ \mbox{mod}\ N)\\
0 & (\mbox{otherwise})
\end{cases}.
\end{align}

{\bf Type 2} makes triplets $(\bm{z}_{i}^{1}, \bm{z}_{i}^{2}, \bm{z}_{j}^{2})$ where $i \not = j$.
With respect to an anchor $\bm{z}_{i}^{1}$, $\bm{z}_{i}^{2}$ is marked as positive and $\bm{z}_{j}^{2}$ is marked as negative.
An example definition of this type is given by
\begin{align}
\alpha_{ij}^{kl} =
\begin{cases}
1 & (k \not = l, i = j)\\
-1 & (k \not = l, i = j-1\ \mbox{mod}\ N)\\
0 & (\mbox{otherwise})
\end{cases}.
\end{align}

{\bf Type 3} makes $(N$$+$$1)$-tuples $(\bm{z}_{i}^{1}, \bm{z}_{1}^{2}, \cdots, \bm{z}_{N}^{2})$.
With respect to an anchor $\bm{z}_{i}^{1}$, $\bm{z}_{i}^{2}$ is marked as positive and all the others are marked as negative.
The definition of this type is given by
\begin{align}
\alpha_{ij}^{kl} =
\begin{cases}
1 & (k < l, i = j)\\
-1 & (k < l, i \not = j)\\
0 & (\mbox{otherwise})
\end{cases}.
\end{align}

{\bf Type 4} makes $2N$-tuples $(\bm{z}_{i}^{1}, \bm{z}_{1}^{2}, \cdots, \bm{z}_{N}^{2},\bm{z}_{1}^{1}, \cdots, \bm{z}_{i-1}^{1},$ $\bm{z}_{i+1}^{1}, \bm{z}_{N}^{1})$.
With respect to an anchor $\bm{z}_{i}^{1}$, $\bm{z}_{i}^{2}$ is marked as positive and all the others are marked as negative.
The definition of this type is given by
\begin{align}
\alpha_{ij}^{kl} =
\begin{cases}
1 & (k \not = l, i = j)\\
0 & (k = l, i = j)\\
-1 & (\mbox{otherwise})
\end{cases}.
\end{align}

\subsection{Representation batch}
Table~\ref{tableB} gives three types of definition
for the representation batch $Z = \{\bm{z}_{i}^{k} : i=1,2,\cdots,N, k=1,2\}$.

{\bf Labeled:}
With labeled samples for supervised learning,
$\bm{z}_{i}^{k}$ denotes the $k$-th representation from class $i$.
A representation $\bm{z}_{i}^{k}$ is defined by
sample representation $\bm{z}_{i}^{k} = f_{\theta} (\bm{x}_{i}^{k'})$
or a statistical representation, such as a mean representation (prototype)
computed from some samples in $B' \subset B$, specifically,
$\bm{z}_{i}^{k} = \frac{1}{|B'|} \sum_{k' \in B'} f_{\theta} (\bm{x}_{i}^{k'})$. Here, $B = \{(\bm{x}_{i}^{k'},y_{i}):i=1,2,\cdots,N, k'=1,2,\cdots,K'\}$ is a mini-batch of labeled samples.

{\bf Labeled+weights:}
This type uses parameters as prototypes,
where $\bm{z}_{i}^{1} = f_{\theta} (\bm{x}_{i}^{1})$ is a representation from class $i$
and $\bm{z}_{i}^{2} = \bm{w}_{i}$ is a weight parameter for class $i$.

{\bf Unlabeled:}
With unlabeled samples for unsupervised learning,
$\bm{z}_{i}^{k}$ denotes the representation of the $i$-th sample with the $k$-th augmentation.
With this type, prototypes can also be introduced
in the same way as prototypes are constructed for labeled samples,
that is, by taking the mean of representations from more than one augmentation function.

\end{document}